\documentclass[11pt]{article}

\usepackage[a4paper,margin=1.2in]{geometry}
\usepackage[semicolon]{natbib}

\usepackage{todonotes}

\usepackage{graphicx}
\usepackage{epstopdf}
\usepackage{url}
\usepackage{caption}
\usepackage{subcaption}
\usepackage{parskip}
\usepackage{needspace}
\usepackage[compact]{titlesec}
\titlelabel{\thetitle. }

\usepackage{color}
\usepackage{amsmath}
\usepackage{amssymb}
\usepackage{amsthm}

\newtheoremstyle{mystyle} 
{5pt} 
{5pt} 
{\itshape} 
{} 
{\bfseries} 
{.} 
{.7em} 
{}

\theoremstyle{mystyle}
\newtheorem{proposition}{Proposition}

\theoremstyle{mystyle}
\newtheorem{lemma}{Lemma}

\newcommand{\cf}{\mathcal{F}} 

\newcommand{\comment}[1]{}

\title{A L\'evy-driven rainfall model with applications to futures pricing}
\date{}
\author{ 
\centering
\begin{tabular}{lll}
Ragnhild C. Noven & Almut E. D. Veraart & Axel Gandy \\[10pt]
\multicolumn{3}{c}{Department of Mathematics, Imperial College London}\end{tabular}
}

\begin{document}

\maketitle

\begin{abstract}
We propose a parsimonious stochastic model for characterising the distributional and temporal properties of rainfall. The model is based on an integrated Ornstein-Uhlenbeck process driven by the Hougaard L\'evy process. We derive properties of this process and propose an extended model which generalises the Ornstein-Uhlenbeck process to the class of continuous-time ARMA (CARMA) processes. The model is illustrated by fitting it to empirical rainfall data on both daily and hourly time scales. It is shown that the model is sufficiently flexible to capture important features of the rainfall process across locations and time scales. Finally we study an application to the pricing of rainfall derivatives which introduces the market price of risk via the Esscher transform. We first give a result specifying the risk-neutral expectation of a general moving average process. Then we illustrate the pricing method by calculating futures prices based on empirical daily rainfall data, where the rainfall process is specified by our model. 
\end{abstract}

\section{Introduction}
A typical rainfall time series has several properties that are difficult to capture in a simple statistical model, including a heavily skewed marginal distribution that is distinctly non-Gaussian, a large proportion of zero values, and frequent large fluctuations. Thus there is a need for specialised models for rainfall which can capture the unique characteristics of this type of process. 
The existing literature on modelling rainfall is large and spread over fields such as hydrology, atmospheric sciences, environmental risk analysis and statistics. 
\citet{Onof2000} classifies the different approaches into four categories: meteorological models based on large sets of differential equations, multi-scale models concerned with the spatial evolution of rainfall, statistical models that capture spatial and temporal trends, and finally stochastic process models that make simple assumptions in order to remain parsimonious. In the following we will focus on models of the last category. 

Many of the early attempts at modelling rainfall use a simple model that represents the rainfall occurrence process as a two-state Markov chain, and models the intensity of rainfall with a Gamma distribution~\citep{Katz1977,chin1977, Woolhiser1982, Coe1982}. This model is easy to interpret and enables the direct use of likelihood methods for fitting. However, it makes several restrictive assumptions on the rainfall process, and may require a high-order Markov chain with many parameters to capture observed temporal dependence. 

There is also a large literature on modelling rainfall for hydrological applications based on a form of the Poisson-cluster model, first developed by \cite{Rodriguez-Iturbe1987} and \cite{Cox1988}. This model is based on a hierarchical structure, with a primary Poisson process controlling the arrival of storms and a secondary process generating cells from each storm, which then deposit rainfall. There have been numerous extensions of the Poisson-cluster model that focus on fitting specific properties of the observed rainfall process. For example, \cite{Cowpertwait1994} considers an extended model that allows for different rainfall cell types, where each cell has a random duration and intensity depending on its type, allowing for the different types of precipitation that are observed in practice.

In terms of fitting specific properties of rainfall that are of interest for hydrological applications, extensions to the Poisson-cluster model generally perform very well. However, as remarked by \citet{Onof2000}, there is always a trade-off between the inclusion of more features and the mathematical tractability of the resulting models. Poisson-cluster based models are usually fitted by the method of moments, which involves matching analytical expressions for properties such as the mean, variance, and proportion of dry intervals to their empirical equivalents. \cite{Chandler1997}, remarks that the method of moments approach ``suffers from the disadvantage that the parameter estimates can vary greatly depending on the properties used in the fitting procedure'', and proposes a spectral estimation method for estimating rainfall models based on point processes.

The contribution of the present paper is twofold. First, we develop a parsimonious and analytically tractable model that captures the distributional features and autocorrelation structure of the observed rainfall time series. 
We relate the model to the framework of L\'evy-driven, continuous-time ARMA (CARMA) processes, and use this connection to develop a suitable fitting method, which is illustrated using empirical rainfall data. The model structure may be interpreted as a non-clustered Poisson model with multiple cell types, cf.~\cite{Cowpertwait1994}. The main benefit of this model 
is its parsimonious formulation based on a stochastic integral, making it suitable for applications where mathematical tractability and fitting methodology are of primary importance. Second, we derive a formula specifying the so-called risk-neutral distribution of a general class of L\'evy-driven stochastic processes, which includes our model as a special case. We then use this result to calculate prices for rainfall futures based on our model.  

Rainfall derivatives were introduced at the Chigago Merchantile Exchange (CME) in 2010, as a recent addition to the class of weather-related products. These products have a large potential market in all economic sectors that depend on favourable weather conditions, such as farming and energy development. The literature on rainfall derivatives pricing is currently rather limited. Because the underlying rainfall cannot be traded directly, the rainfall derivative market is incomplete, and thus there is no single fixed price for the derivative. Due to this incompleteness, there are several distinct methods that can be used for rainfall derivative pricing. It appears that most current approaches rely on either the utility indifference approach or risk-neutral pricing using the Esscher transform.

The utility indifference approach is used in \citet{Carmona2005}. In this paper a modification of the Poisson-cluster model is considered, which makes the rainfall intensity a Markov jump process, thus enabling maximum likelihood estimation. This modification relies on the assumption that the data used, though inevitably discrete, approximates continuous-time observations of the rainfall intensity. \citet{Leobacher2009} also use the utility indifference approach for pricing hypothetical rainfall derivatives in Kenya, based on a Markov-Gamma model with seasonality.

A natural choice for pricing based on L\'evy process models is the Esscher transform \citep{Esscher1932}, as it is structure-preserving \citep{Esche2005} and moreover gives rise to a minimal entropy martingale measure \citep{Frittelli2000}. \citet{LopezCabrera2013} use the Esscher transform for pricing based on a version of the daily rainfall model by \cite{Wilks1998}. They fit simulated monthly rainfall totals to a normal inverse Gaussian distribution, and use the Esscher transform to obtain a risk-neutral distribution. \citet{Benth2013} also use the Esscher transform for pricing, but base their underlying rainfall model on an independent increment process.

In contrast to the Markov-Gamma and independent increment models that have been used for pricing rainfall derivatives, our proposed model has the advantage of not making assumptions about temporal independence or Markovianity of the rainfall process increments. Furthermore, the model and the fitting method used are based on the assumption that the available data represents accumulated rainfall, i.e.~the instantaneous intensity cannot be directly observed, which is typically the case for applications. By allowing for temporal dependence and considering CARMA processes of arbitrary order, we obtain a model with a flexible autocorrelation structure, which is particularly relevant for data on finer time scales. This flexibility is illustrated by fitting the model to hourly rainfall data. We also compare our model to that given in \cite{Wilks1998}, which was used in \cite{LopezCabrera2013} for pricing rainfall futures based on daily data from Detroit. 
 
This paper is structured as follows: Section 2 discusses characteristic features of the rainfall process in light of data from different locations and time scales. Section 3 presents the rainfall model and shows how it fits into the continuous-time ARMA (CARMA) model framework. Section 4 gives details on the fitting method, and Section 5 investigates the model performance using empirical data. In Section 6 we derive a method for pricing rainfall derivatives based on our model.   

\section{Characteristics of observed rainfall} \label{Empirical}
In this section we motivate the structure of our model by illustrating some of the characterising features of rainfall time series. We base this illustration on two data sets which will be used throughout this paper: the first consists of hourly accumulated rainfall amounts at Heathrow (UK) over the years 1980-2012, provided by the \cite{MIDAS}. The second data set gives daily accumulated rainfall amounts in Detroit (US) over the years 1980-2010, provided by Bloomberg Professional Service. 

\begin{figure*}[tbp]
	\centering
	\includegraphics[width=0.9\textwidth]{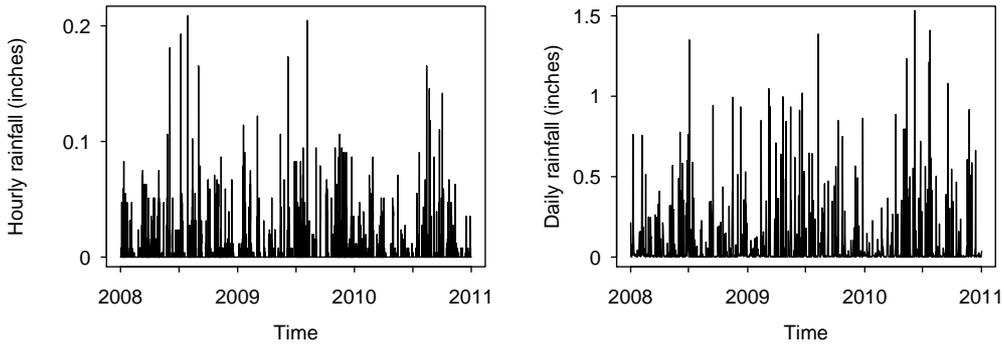}
\caption{Time series of empirical rainfall over three years for Heathrow (left) and Detroit (right).}
	\label{detroit_heathrow_3yrs}
\end{figure*}

Figure \ref{detroit_heathrow_3yrs} shows the rainfall time series for both locations over the years 2008-2010. These plots illustrate that the rainfall process is subject to sudden transitions between periods with little or no rain and periods of higher intensity, causing the large spikes in the graph. 

There is also a large amount of zero values in both data sets, as shown in Table \ref{zerotable}. Because the data is rounded to the nearest unit of measurement ($0.1$mm for the Heathrow data and $3/100$ inch for the Detroit data), these zero values represent all data points with a value smaller than this unit. The proportion of zero values is dependent on the measurement time scale, with fewer zero values for the daily data. This is because periods with no rainfall must persist throughout the measurement time interval in order to induce a value of zero in the data.
\begin{table}[tbp] 
\centering
\caption{Percentage of rainfall measurements equal to zero.}
\label{zerotable}
\begin{tabular}{llll}
\hline
Location & Heathrow & Heathrow & Detroit \\
\hline
Time scale & Hourly & Daily & Daily \\
Zero values (\%) &  91.27 &  53.15 & 48.14 \\ 
\hline
\end{tabular}
\end{table}

\begin{figure*}[tbp]
	\centering
		\includegraphics[width=0.9\textwidth]{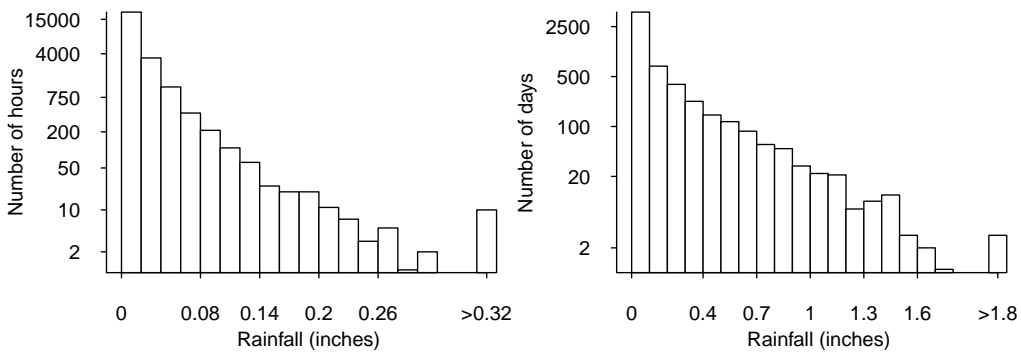}
		\caption{Frequency plots of non-zero rainfall for Heathrow (left) and Detroit (right), with frequencies on a log scale and the upper end of the range grouped together.}
	\label{fig:detroit_heathrow_nonzero_hist}
\end{figure*}

Figure \ref{fig:detroit_heathrow_nonzero_hist} shows frequency plots of the non-zero (i.e.~positive-valued) data, which illustrates the non-normality and skewness of the empirical distributions. In general, hourly data has more pronounced skewness, for Heathrow the coefficient is 15.87, compared with 4.18 when the data is aggregated to the daily time scale. This fits with the general observation that measuring accumulated rainfall on larger time scales has a smoothing effect, which makes characterising features such as large skewness and frequent zero values less evident.  

\begin{figure*}[tbp]
	\centering
	\includegraphics[width=0.9\textwidth]{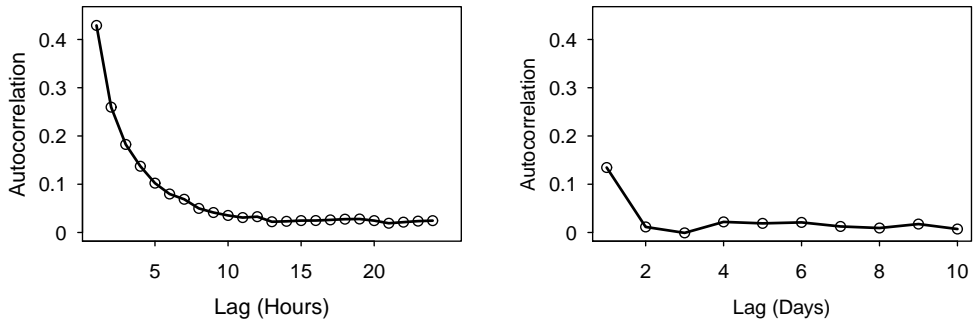}
        \caption{Empirical  rainfall autocorrelation functions for Heathrow (left) and Detroit (right).}
	\label{detroit_heathrow_empirical_acf}
\end{figure*}

The empirical autocorrelation functions of the Heathrow and Detroit rainfall time series are shown in Figure \ref{detroit_heathrow_empirical_acf}. For the hourly Heathrow data there is clearly a non-trivial autocorrelation part which decays smoothly up to lag 10. For the daily Detroit data the autocorrelation function (ACF) decays steeply after lag 1, indicating that there is less relevant time-dependence in this rainfall process, as one would expect from the daily time scale. As will be seen in Section \ref{sfitting}, these differences in the autocorrelation structure lead us to fit models of different orders to the two data sets. 
 
\section{Rainfall Model} \label{model}
In this section we present the basic structure of our model and relate its properties to the observed rainfall dynamics. Based on this we construct a model extension where the rainfall intensity process belongs to the class of continuous-time ARMA (CARMA) models.   

\subsection{Primary model structure} \label{primary}
 We propose to model the accumulated rainfall $R$ by 
\begin{align} \label{modeldef}
R(t_i)-R(t_{i-1})=S(t_{i-1})(Y(t_i)-Y(t_{i-1})),
\end{align} where $0=t_0<t_1<\ldots<t_n$ are discrete measurement times such that $t_i-t_{i-1}=\delta$, and $S$ is a deterministic seasonal component, {which will be discussed in detail in Section \ref{seasonality}. We define the continuous-time stochastic process $(Y(t))_{t \geq 0}$ as the integral of a L\'evy-driven Ornstein-Uhlenbeck (OU) process $(X(t))_{t \geq 0}$ \citep{Barndorff-Nielsen2001}, i.e.
\begin{align} 
\begin{split}
Y(t)&=\int_0^t X(s) \, ds, \label{xeqn} \\
X(s)&=X(0) e^{-\lambda s}+\int_0^s e^{-\lambda(s-v)} \, dL(v),
\end{split}
\end{align} 
where $\lambda$ is a positive parameter and $(L(t))_{t \geq 0}$ is a L\'evy subordinator. We interpret $X(s)$ as the instantaneous rainfall intensity at time $s$, and so $Y(t)$ measures the accumulated rainfall over the time interval $[0,t]$ - up to the seasonal adjustment given by $S$. We let $X(0)$ be a random variable which is independent of $(L(t))_{t \geq 0}$ and has distribution
\[X(0)\stackrel{d}{=}\int_0^\infty e^{-\lambda v} \, dL(v),\] 
making the resulting OU process $X$ strictly stationary. 

Using the Fubini theorem for stochastic integrals, we can exchange the order of integration in the definition of $Y$ and obtain a simpler expression in terms of a single stochastic integral: 
\begin{align} \label{yeqn}
Y(t)=X(0)\left(\frac{1-e^{-\lambda t}}{\lambda}\right)+\int_0^t \frac{1-e^{-\lambda(t-v)}}{\lambda} \, dL(v). 
\end{align}

For our particular rainfall model we let $(L(t))_{t \geq 0}$ be a compound Poisson process with Gamma-distributed jumps, called the Hougaard process \citep{Lee1993,Grigelionis2011}. This means that $L$ is a pure-jump L\'evy process, more specifically a subordinator. The choice of a pure-jump L\'evy process is motivated by the intermittent behaviour of the observed rainfall process, in particular the abrupt switches from exact zero to large positive values, which are modelled by jumps in the driving process $L$.  

The marginal distribution of $L(1)$ is a member of the Tweedie distribution family~\citep{Jorgensen1997}, which was used by \citet{Dunn2004} to model the monthly rainfall in Australia. In the following we will parameterise $L(1)$ as a Tweedie random variable, which has parameters $(\mu,\rho,\kappa)$ such that
\begin{align} \label{lspec}
\begin{split}
\mbox{E}(L(1))&=\mu,\\
\mbox{Var}(L(1))&=\rho \mu^\kappa.
\end{split}
\end{align} 

The stochastic process $X$ defined in \eqref{xeqn}, which represents the rainfall intensity, has an interpretation in terms of the physical dynamics of the rainfall process. In this interpretation the jumps of the driving process $L$ represent the arrival of storm events, generating a jump in the intensity of random size. As the storm dissipates, this intensity decays smoothly towards zero at a rate determined by the parameter $\lambda$ in the OU process. 

By using this interpretation we see the paralell to the Poisson-cluster models discussed in the introduction, which are based on the idea of storms arriving according to a Poisson process. More specifically, the basic model presented in this section can be interpreted as a non-clustered Poisson model, i.e.~the special case where each storm has only one cell. At each storm arrival, the current intensity increases by a random, Gamma-distributed amount, and then decays exponentially from the increased level.

A similar approach was considered in \cite{Samuel1999}, under the name of ``Poisson Decaying pulse model'', corresponding to a non-clustered Poisson model where each cell has an exponentially decreasing intensity, with the addition that each cell has a random lifetime after which its intensity is set to zero. An attempt was made to fit this model using the spectral method \citep{Chandler1997}, however it was found that there is no unique solution to this estimation problem.  

Coming back to our model structure, it is clear from $\eqref{modeldef}$ that the discrete-time process $(\triangle Y(t_i))_{i=1,\ldots,n}$ given by
\[\triangle Y(t_i):=Y(t_i)-Y(t_{i-1})\]
should have features resembling those of deseasonalised empirical rainfall. As shown in Section \ref{results}, Figures \ref{heathrow_double_hist} and \ref{detroit_triple_hist}, the empirical marginal distribution of our rainfall data is well approximated by the marginal distribution of $\triangle Y$. In fact, this was the motivation for specifying the driving process $L$ to be the Hougaard process. 

When it comes to approximating the empirical autocorrelation structure, the present model is rather limited, because the autocovariance function $C_{\triangle Y}(h)$ is restricted to take the form of an exponential decay. Specifically, we get that
\begin{align} 
\begin{split}
&C_{\triangle Y}(0)= \frac{\rho \mu^\kappa}{\lambda^3} (e^{-\lambda \delta}+\lambda \delta-1) \label{yacf}, \\
&C_{\triangle Y}(h)= \frac{\rho \mu^\kappa}{2 \lambda^3} (e^{-\lambda \delta}-1)^2 e^{-\lambda(h-1)\delta}, \quad h \geq 1.
\end{split}
\end{align}

As illustrated in Figure \ref{detroit_heathrow_empirical_acf}, the empirical autocovariance functions do not necessarily take such a simple form. This restrictive form of the autocovariance function motivates the following extension of the model. 

\subsection{Extension to CARMA process} \label{CARMA}
In this subsection we consider an extension of our model which admits a more flexible autocovariance structure. This extension is based on generalising the Ornstein-Uhlenbeck process $X$ in \eqref{xeqn} to a continuous-time ARMA (CARMA) process. We first give a brief overview of the construction of L\'evy-driven CARMA processes, and then show how the extension of $X$ is obtained. 

\subsubsection{CARMA processes}
A CARMA processes is a continuous-time analogue of the discrete-time ARMA process. Here we will consider L\'evy-driven CARMA processes \citep{Brockwell2001,Brockwell2009}. To illustrate the correspondence to the discrete-time setting, we start by considering the ARMA$(p,q)$ process $(V_n)$ defined by the difference equation
\[a(B)V=b(B)L, \] 
where $B$ is the backward shift operator, $L$ is a white noise sequence and $a,b$ are polynomials given by
\begin{align*}
a(x)&=x^p+a_1x^{p-1}+\ldots+a_p,\\
b(x)&=b_0+b_1x+\ldots+b_qz^q.
\end{align*}
We can consider formally replacing $B$ with the differential operator $D$ to obtain a stochastic differential equation (SDE) for the CARMA$(p,q)$ process $V$ driven by the process $L$.  

This SDE will contain expressions of the form $D^k L$, which may not be well-defined. Therefore it is customary to consider an equivalent definition of CARMA processes via the state-space representation. This representation defines the observation and state equations 
\begin{align*}
&V(t)=\mathbf{b}^T \mathbf{Z}(t),\\
&d\mathbf{Z}(t)-A\mathbf{Z}(t)dt=\mathbf{e} \,dL(t),
\end{align*}
where $\mathbf{b}$ is the vector of coefficients of $b(x)$, $A$ is the matrix 
\[
\left(
\begin{array}[]{lllll}
0&1&0&\ldots&0\\
0&0&1&\ldots&0\\
\vdots&\vdots&\vdots&\vdots&\vdots\\
-a_p&-a_{p-1}&-a_{p-2}&\ldots&-a_1\\
\end{array}
\right),
\]
where $a_i$ is the ith coefficient of $a(x)$, $\mathbf{e}$ is the pth unit vector and $L$ is the driving L\'evy process, where $E \max(0,\log |L(1)|)<\infty$. Provided all eigenvalues of $A$ have negative real parts, the SDE for $Z$ can be solved to give the following expression for $(V(t))_{t \geq 0}$:
\begin{align} \label{kernel_carma}
V(t)=\mathbf{b}^T e^{At} \mathbf{e} V(0)+\int_0^t \mathbf{b}^T e^{A(t-u)} \mathbf{e} \, dL(u),
\end{align} 
where $V$ is strictly stationary. 
We can now obtain a representation of the CARMA process $V$ that extends the L\'evy-driven OU process \citep{Brockwell2004}. Assume that $A$ has distinct eigenvalues $\{\alpha_i\}$ (equivalently, that the polynomial $a(x)$ has distinct roots $\{\alpha_i\}$), with corresponding eigenvectors 
\[(1,\alpha_i,\ldots,\alpha_i^{p-1}).\]
Then we can obtain the spectral expansion
\[\mathbf{b}^T e^{A(t-u)} \mathbf{e}=\sum_{i=1}^p \frac{b(\alpha_i)}{a'(\alpha_i)} e^{\alpha_i (t-u)}.\]

Substituting this expansion into \eqref{kernel_carma} gives 
\begin{align} \label{expcarma}
V(t)= V(0)\left(\sum_{i=1}^p \frac{b(\alpha_i)}{a'(\alpha_i)} e^{\alpha_i t} \right)+\int_0^t \left(\sum_{i=1}^p \frac{b(\alpha_i)}{a'(\alpha_i)} e^{\alpha_i (t-u)}\right) \, dL(u), \; t \geq 0.
\end{align} 

If we now set $p=1$ and let $a(z)=z-\lambda, b(z)=1$,  we recover the OU process $X$ given in \eqref{xeqn}, with $\lambda=-\alpha_1$. Thus CARMA processes generalise Ornstein-Uhlenbeck processes, and this motivates the extended model described in the following.  

\subsubsection{Extended model} \label{extended}
We define the extended model of order $p$ by
\[R(t_i)-R(t_{i-1})=S(t_{i-1})(Y(t_i)-Y(t_{i-1})),\]
where 
\begin{align} 
&Y(t)=\int_0^t X(s) \, ds, \label{yext}\\
&X(s)=\left(\sum_{i=1}^p w_i  X_i(0)  e^{-\lambda_i s}\right)+\int_0^s \left( \sum_{i=1}^p w_i e^{-\lambda_i(s-v)}\right) \, dL(v)  := \sum_{i=1}^p w_i X_i(s) \label{xmixt},
\end{align}
with $\sum_i w_i=1$. Here $X$ is a CARMA process of the form given in \eqref{expcarma}, where the  coefficients of the polynomial $b$ can be found by solving $b(-\lambda_i)/a'(-\lambda_i)$ for $w_i$, with $b_q=1$. In the following we will assume that $b$ has order $q=p-1$, making $X$ a CARMA$(p,p-1)$ process. The requirement $q=p-1$ is necessary for obtaining the implied ARMA process representation \citep{Brockwell2013} used in the fitting method described in Section \ref{impliedarma}. 

Similarly to the OU case, each $X_i(0)$ is chosen to be independent of $(L_t)_{t \geq 0}$, with
\[\sum_{i=1}^p w_i  X_i(0)  e^{-\lambda_i s} \stackrel{d}{=}\int_0^\infty \left(\sum_{i=1}^p w_i  e^{-\lambda_i (s+v)}\right) \, dL(v).\]
This extension of $X$ can also be seen as a mixture of \emph{dependent} OU processes $X_i$, driven by the same subordinator $L$.

Following the interpretation given in the previous subsection, storms arrive in a Poisson process, generating a jump in the intensity which is Gamma distributed. In the extended model, the intensity until the next arrival is given by a weighted sum of $p$ intensity processes, which decay from the \emph{same} initial level at different rates $\lambda_i$. This could be taken to mean that a typical storm has $p$ components whose intensity dissipates at different rates.

As we will see in equation \eqref{mixtcovar}, the autocovariance function $C_{\triangle Y}$ with $Y$ defined in \eqref{yext} is a mixture of exponential decays with separate rates $\lambda_i$. Hence we can get more complex autocovariance structures by increasing the order $p$ of the CARMA process $X$.   

\section{Fitting procedure} \label{sfitting}
In this section we discuss a fitting approach for the extended model defined in Section \ref{extended}. The fitting is done in three parts, firstly the deterministic seasonality function $S$ is estimated, then we estimate the autocovariance parameters via the CARMA representation, and finally we find moment-based estimates of the driving L\'evy process parameters.

\subsection{Seasonality function} \label{seasonality}
The multiplicative seasonality function $S$ is estimated in an ad hoc fashion by fitting a truncated Fourier series with an annual period to the empirical mean of each month, specifically we have 
\begin{align} \label{fseries}
S(t)=\frac{a_0}{2}+\sum_{i=1}^n a_i \cos(2 \pi i t/12)+b_i \sin(2 \pi i t/12),
\end{align}
for time $t$ on a monthly scale, where $a_i,b_i$ are the fitted parameters. The Fourier series was fitted by considering a linear model where the responses are the empirical monthly means and the covariates are the corresponding values of the $\sin$ and $\cos$ terms. The order of truncation $n$ was then chosen to minimise the AIC of this linear model, which occurs at order $n=2$, giving a total of $5$ parameters for the seasonality function.

Using a multiplicative seasonality function has the advantage of enabling straightforward fitting of a single integrated CARMA process $Y$ to all the available data. However, it does not account for seasonal variations in the autocorrelation structure or higher-order moments of $\triangle Y$. An alternative approach would be to first fit the model separately to data from each month, in order to detect any significant seasonal changes in the parameters. The monthly parameter values could then be allowed to vary between months according to a suitably chosen, smoothly varying function. Using this setup, the model can be fitted by simultaneously minimising the squared prediction errors and moment differences for the individual monthly models.

In the present paper we will only consider the multiplicative seasonality function, and focus on fitting the integrated CARMA process to the deseasonalised data. After fitting the seasonality function $S$ using a truncated Fourier series as described above, we rewrite \eqref{modeldef} as
\begin{align*} 
\frac{R(t_i)-R(t_{i-1})}{S(t_{i-1})}=Y(t_i)-Y(t_{i-1}),
\end{align*}
which shows that we can fit $\triangle Y$ to the discrete observations $\triangle R/S$. Thus in the following we will only consider fitting the model given by $\triangle Y$.

\subsection{Autocovariance structure} \label{impliedarma}
We now show how to use the CARMA representation of the process $X$ to develop a fitting method for the parameters $\{\lambda_i, w_i\}$. This approach relies on Theorem 2 in \citet{Brockwell2013}, which states that under certain conditions\footnote{The conditions are as follows:  $a$ and $b$ have no common zeroes, the roots of $a$ have multiplicity 1, and $\mbox{Im}(\lambda_i) \in (-\frac{\pi}{\delta},\frac{\pi}{\delta})$.} on the polynomials $a$ and $b$, we have that for a causal and invertible CARMA$(p,p-1)$ process $V$, the discrete process $I_n^\Delta$ given by
\[I_n^\Delta=\int_{(n-1)\Delta}^{n \Delta} V(s) ds,\]
is a weak ARMA$(p,p)$ process. This implied ARMA process takes the form
\[\phi(B)I_n^\Delta=\theta(B)\epsilon_n,\]
where $\{\epsilon_i\}$ is a weak white noise sequence, i.e.~the terms are uncorrelated but possibly dependent. Here the parameters of the process $L$ driving the CARMA process $V$ only affect the sequence $\{\epsilon_i\}$, not the polynomials $\phi$ and $\theta$. Furthermore, the theorem also states that there is a one-to-one correspondence between the coefficients $\{w_i, \lambda_i\}$ of $V$ and the coefficients $(\phi_i,\theta_i)$ of the corresponding ARMA$(p,p)$ process. 

By using the CARMA$(p,p-1)$ representation of $X$ as defined in \eqref{xmixt}, we can write 
\[\triangle Y(t_i)=\int_{t_{i-1}}^{t_i} X(s) \, ds= I_{t_i/\delta}^\delta.\]
Hence the observed increments of $Y$ can be seen as observations from the implied weak ARMA process.   

We can also obtain the autocovariance of $\triangle Y$ from the integrated CARMA representation \citep[Corollary 2]{Brockwell2013}:
\begin{align} \label{mixtcovar}
\begin{split}
C_{\triangle_Y}(0)&=\sum_{\lambda_i}  2 \beta(\lambda_i )\lambda_i^{-2}(e^{-\lambda \delta}-1+\lambda_i \delta),\\
C_{\triangle_Y}(h) &= \sum_{\lambda_i}  \beta(\lambda_i ) \lambda_i^{-2} (e^{-\lambda_i \delta}-1)^2 e^{-\lambda_i(h-1)\delta}, \quad h \geq 1, \\
\beta(\lambda_i)&=\sigma^2 \frac{b(-\lambda_i)b(\lambda_i)}{a'(-\lambda_i)a(\lambda_i)},
\end{split}
\end{align}
where $\sigma^2$ is the variance of the driving process increment $L(1)$, and $a,b$ are the polynomials in the CARMA representation of $X$. For the Hougaard process we have $\sigma^2=\rho \mu^\kappa$. 

We now follow~\citet{Brockwell2013} in estimating $\{w_i,\lambda_i\}$ by minimising the weighted sum of the one-step prediction errors of the implied ARMA$(p,p)$ process, which is equivalent to minimising with respect to $\{w_i, \lambda_i\}$ due to the one-to-one correspondence. Initial values for the parameters in the minimisation can be obtained by setting the values of the autocovariance function of $\triangle Y$ for the first few lags equal to the corresponding empirical values. The estimation procedure based on minimising the prediction errors is shown to be strongly consistent by \citet{Brockwell2013}. 

We will use a CARMA$(1,0)$ model for the intensity process $X$ corresponding to the Detroit rainfall data, and a CARMA$(2,1)$ model for $X$ corresponding to the Heathrow data. These orders are chosen to be as low as possible while ensuring that the ACF of the fitted model can adequately replicate the shape of the empirical ACF. 

If the order $p$ of the model is chosen too high, some of the weight parameters $w_i$ may have estimates equal to $0$, meaning that the process $X_i$ has no influence on $X$. Thus the fitted model is equivalent to specifying $X$ with a lower order $p'<p$.  This was found to be the case for the Detroit data when using the model with $p=2$, resulting in the estimates $\hat{w_1} \approx 1, \hat{w_2} \approx 0$. This motivates the use of a CARMA$(1,0)$ process for the Detroit data. In general we expect that higher order models are more suitable for high-frequency data, which has more significant dependence structure.

For the CARMA$(1,0)$ model representing the Detroit data we have the injective mapping $\lambda \rightarrow (\phi,\theta)$ given by
\begin{align*}
\phi(\lambda)&=e^{-\lambda},\\
\theta(\lambda)&=-r-\sqrt{r^2-1},\\
r&=\frac{1-\lambda-e^{-2 \lambda}(1+\lambda)}{1-2 \lambda e^{-\lambda}-e^{-2 \lambda}}.
\end{align*}

For the CARMA$(2,1)$ model, the mapping between $(w_1,w_2,\lambda_1,\lambda_2)$} and $(\phi_1,\phi_2,\theta_1,\theta_2)$ is found by numerically solving for the autocovariance function of the implied ARMA process, under the constraint $w_1+w_2=1$. 

\subsection{Driving L\'evy process}
Having estimated $\{w_i,\lambda_i\}$ it remains to estimate the parameters $(\mu,\rho,\kappa)$ of the driving process $L$. The parameter estimation is done by the method of moments applied to the process $\triangle Y$, which can be written as
\begin{align} \label{int_form}
\begin{split}
\triangle Y(t_i)=& \int_0^\infty  \sum_{k=1}^p \left(\frac{e^{-\lambda_k t_{i-1}}-e^{-\lambda_k t_{i}}}{\lambda_k} \right)  e^{-\lambda_k v} \, dL^{\ast}(v)+\int_0^{t_{i-1}} \sum_{k=1}^p \left(\frac{e^{-\lambda_k (t_{i-1}-v)}-e^{-\lambda_k (t_{i}-v)}}{\lambda_k} \right) \, dL(v)\\
+&\int_{t_{i-1}}^{t_i} \sum_{k=1}^p \left(\frac{1-e^{-\lambda_k(t_i-v)}}{\lambda_k}\right) \, dL(v), 
\end{split}
\end{align}
where the processes $L,L^{\ast}$ are independent and have the same characteristic triplet, given by $(0,0,\nu(\cdot))$, corresponding to the Hougaard process.

Using the above representation we can obtain an analytic expression for the characteristic function of $\triangle Y$, as shown in the Appendix. Using this expression, we can find the moments of $\triangle Y$ and thus fit the driving L\'evy process. It is readily shown that  
\[\mbox{E}(\triangle Y)=\sum_{k=1}^p \frac{\mu \delta}{\lambda_k},\]
and furthermore the variance of $\triangle Y$ has been given in \eqref{mixtcovar}. The third moment is calculated numerically from the characteristic function.

We now replace the autocovariance parameters $\{w_i,\lambda_i\}$ in the expressions for the moments with their least-squares estimates. Comparing the theoretical moments to those of the observed increments $\triangle R/S$ gives three equations with unknowns $(\mu,\rho,\kappa)$, which can be solved to obtain estimates for these parameters. 

To obtain confidence intervals for the estimated parameters we use the block bootstrap method \citep{Politis1994,kunsch1989} to resample from the empirical distribution under the assumption of dependent data. In this resampling the block size has a geometric distribution with a specified mean value, which ensures stationarity of the resampled sequence.  

In order to choose the mean block size we first generated a sample of model simulations with the parameters held fixed at their estimated values. Then we applied the block bootstrap with fixed mean block size to the simulated data sets to obtain a bootstrap confidence interval for each simulation. We repeated this procedure for several choices of the mean block size, and chose the one that optimised the coverage rate of the confidence intervals relative to the $95\%$ nominal rate.

After obtaining a bootstrap sample as described above, we calculated 95\% confidence intervals by taking the lower and upper bounds to equal, respectively, the 2.5 and 97.5 percentiles of the bootstrap sample. Tables \ref{hpars} and \ref{dpars} show the estimated parameters and confidence intervals for the Heathrow and Detroit data. 

The parameter estimates for the hourly Heathrow rainfall data show that the autocorrelation structure consists of one quickly decaying component with rate $\lambda_1=4.79$, and one slowly decaying component with rate $\lambda_2=0.31$. These components could be taken to represent different types of storms, perhaps corresponding to the ``light'' and ``heavy'' rainfall cell types considered in \cite{Cowpertwait1994}.

The confidence intervals indicate that for both data sets the estimates of $\lambda_1$ and $\mu$ are quite variable. However, the bootstrap estimates of $w_1,\lambda_1$ and $\lambda_2$ could be affected by seasonal variation in the autocorrelation structure, which is not accounted for by dividing out the seasonality component $S$, as discussed in Section \ref{seasonality}.

\begin{table}[tbp]
\centering
\caption{Estimated parameters and confidence intervals (CI) for Heathrow rainfall data, where ``in'' denotes a unit of inches.}
\label{hpars}
		\begin{tabular}{lr@{}lr@{}lr@{}l}
		\hline \\[-9pt] 
		Parameter  &&$w_1$ &&$\lambda_1$ (\small h$^{-1}$) && $\lambda_2$ (\small h$^{-1}$) \\[1pt]
		Estimated value &$0$&$.92$ &$4$&$.79$ &$0$&$.31$ \\[1pt]
		$95$\% CI &$( 0.88$&$, 0.95)$ &$(3.58$&$, 7.86)$ &$(0.28$&$,0.36)$ \\
		\hline
		&&&&&& \\
		\hline \\[-9pt]
		Parameter &&$\mu$ (\small in) &&$\rho$ (\small in$^{2-\kappa}$) &&$\kappa$ \\[1pt]
		Estimated value &$2$&$.15$ &$143$&$.01$ &$1$&$.85$ \\[1pt]
		$95$\% CI &$(1.64$&$,3.48)$ &$(130.63$&$,158.53)$ &$(1.81$&$,1.92)$ \\
		\hline
		\end{tabular}
\end{table}

\begin{table}[tbp]
\centering
\caption{Estimated parameters and confidence intervals (CI) for Detroit rainfall data, where ``in'' denotes a unit of inches.}
\label{dpars}
		\begin{tabular}{lr@{}lr@{}lr@{}lr@{}l}
		\hline \\[-9pt]
		Parameter &&$\lambda$ (\small d$^{-1})$ &&$\mu$ (\small in) &&$\rho$ (\small in$^{2-\kappa})$ &&$\kappa$\\ [1pt]
		Estimated value &$4$&$.54$ &$4$&$.55$ &$14$&$.85$ &$1$&$.62$ \\ [1pt]
		$95$\% CI &$(4.06$&$,5.25)$ &$(4.07$&$,5.26)$ &$(14.71$&$,14.99)$ &$(1.60$&$,1.64)$ \\
		\hline
		\end{tabular}
\end{table}

\section{Assessing model performance} \label{results}
To assess the fit of the model we first compare several properties of the simulations from the fitted model to the corresponding empirical properties, specifically considering the overall marginal distribution, autocorrelation function and zero proportion of the rainfall process. In the last subsection we compare properties that are important for the pricing application in Section \ref{pricing} over individual months.

A simulation from the process $Y$ can be obtained by using the compound Poisson process representation of the driving process $L$, which gives an expression for $Y$ as a weighted sum of the jumps of $L$. Multiplying by the seasonality function $S$ then gives a simulation from the full model for the accumulated rainfall increments $\triangle R$. 

\subsection{Marginal distribution} \label{marginal}
\begin{figure*}[tb] 
	\centering
	\includegraphics[width=0.8\textwidth]{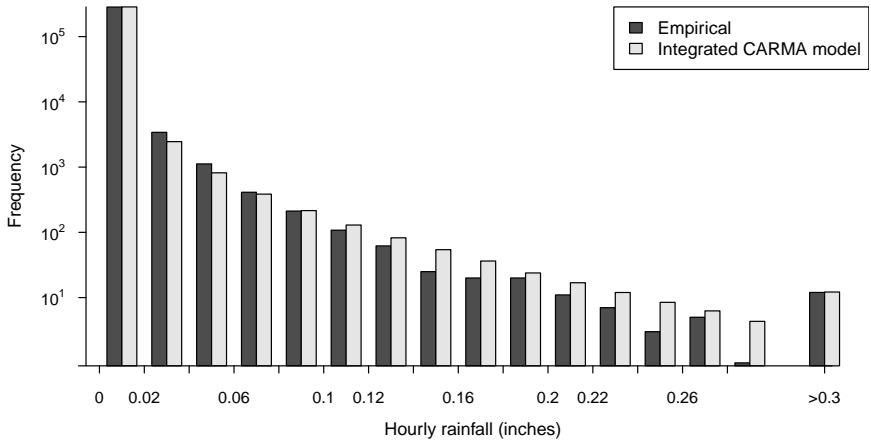}
	\caption{Frequency plots of empirical hourly Heathrow rainfall and simulations from the integrated CARMA model, with frequencies on a log scale and the upper end of the range grouped together.}
	\label{heathrow_double_hist}
\end{figure*}

\begin{figure*}[tb] 
	\centering
	\includegraphics[width=0.8\textwidth]{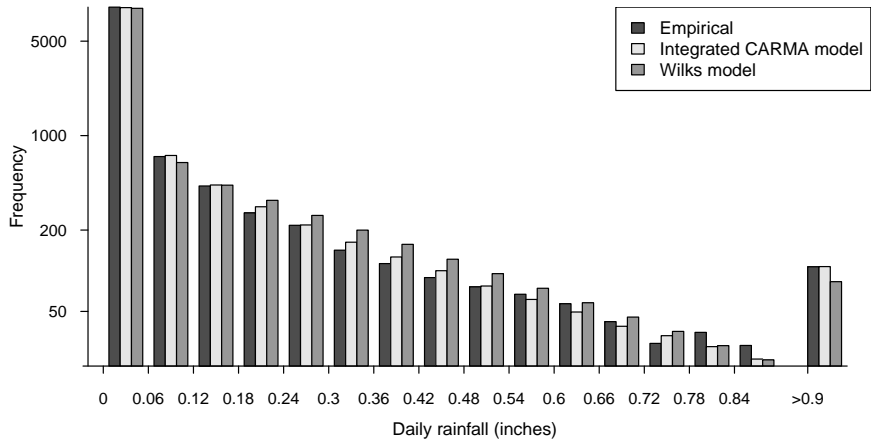}
	\caption{Frequency plots of empirical daily Detroit rainfall and simulations from the integrated CARMA model and the Wilks model, with frequencies on a log scale and the upper end of the range grouped together.}
	\label{detroit_triple_hist}
\end{figure*}

Figures \ref{heathrow_double_hist} and \ref{detroit_triple_hist} show frequency plots of the empirical rainfall time series, together with an estimate of the corresponding model-based frequencies. This estimate is obtained by averaging the frequencies over 100 simulations for the hourly Heathrow model, and 500 simulations for the daily Detroit model, which keeps the computational effort reasonable. Each simulated time series uses the parameter estimates given in Tables \ref{hpars} and \ref{dpars}, and has the same length as the empirical time series. For the daily Detroit data we also include the frequencies averaged over 500 simulations from the model given in \cite{Wilks1998}, which was fitted to the empirical data. This model was used in \cite{LopezCabrera2013} as the basis for a pricing method, and so in view of the application in Section \ref{pricing} it is a natural choice for comparison. 

We see that on both time scales the model manages to capture the characteristic shape of the rainfall distribution quite well. For the daily Detroit data the fit is somewhat better than that of the model given in \cite{Wilks1998}, especially in the lower part of the range where the majority of the data is found.

\begin{figure*}[tb] 
   \centering
   \includegraphics[width=0.8\textwidth]{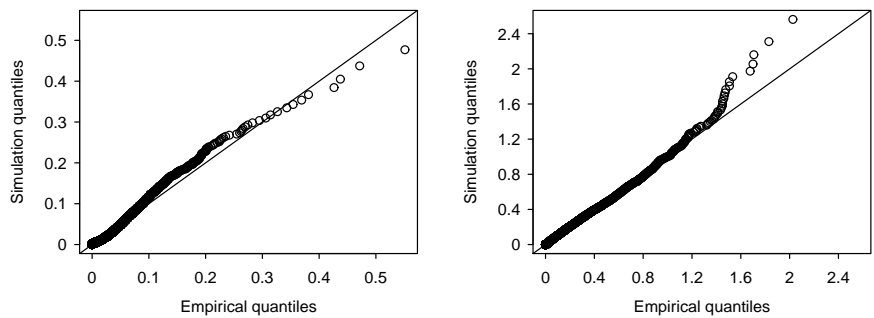}
   \caption{QQ-plots of empirical and simulated rainfall for Heathrow (left) and Detroit (right).}
   \label{qqplots}
 \end{figure*}

Figure \ref{qqplots} shows QQ-plots comparing empirical quantiles to simulation quantiles, where the latter are obtained by combining the data from the respective collections of 100 and 500 simulations for Heathrow and Detroit. These plots confirm the goodness-of-fit in the lower part of the range, however there are some deviations in the extreme quantiles, especially for the daily Detroit model.  

\subsection{Autocorrelation structure} \label{acf_struct}
\begin{figure*}[tbp]
	\centering
	\includegraphics[width=0.8\textwidth]{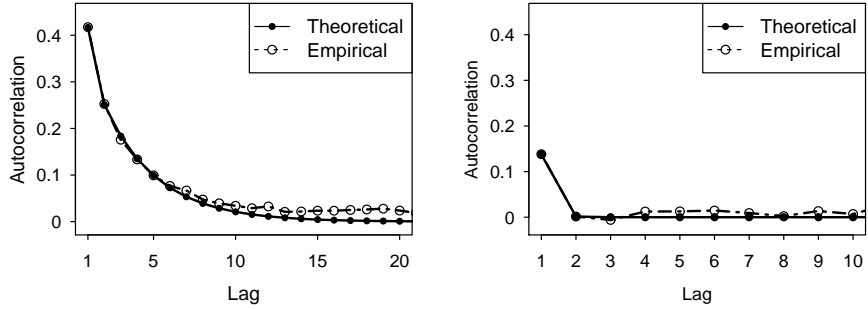}
	\caption{Theoretical and empirical ACF for Heathrow (left) and Detroit (right) rainfall.}
	\label{fig:detroit_heathrow_acf}
\end{figure*}

Figure \ref{fig:detroit_heathrow_acf} shows the autocorrelation function of the deseasonalised data along with the theoretical ACF of the fitted model for both time scales. For the hourly Heathrow data the gradual decay is captured very well by  the fitted ACF from the CARMA$(2,1)$ model, especially below lag 10. For higher lags it appears that the empirical ACF decays somewhat more slowly than the fitted ACF, however the difference is very small, and may be intepreted as noise, or as an effect of the deseasonalisation. If there is evidence of long-range dependence in the data, this could potentially be modelled by using a superposition of OU processes \citep{Barndorff-Nielsen2001a}, although this was not the case for the Heathrow data.

The right panel in Figure \ref{fig:detroit_heathrow_acf} shows that the fitted ACF from the CARMA$(1,0)$ model is very similar to the empirical ACF, in both cases the autocorrelation decays to zero almost immediately. Thus we see that our model manages to capture the autocorrelation structure of the rainfall process accurately for both hourly and daily time scales. 

\subsection{Zero proportion}

Table \ref{zero} shows the proportion of zero values in the empirical data, together with the corresponding zero proportion averaged over 20 simulations. As mentioned in Section \ref{Empirical}, the zero values in the empirical data come from data points with a value below the measurement threshold, and we performed the same rounding for the simulated data to get the \emph{implied} zero proportion shown. We see that the simulated time series for the hourly rainfall have very similar zero proportions to the empirical data, whereas the daily rainfall simulations are somewhat less accurate, they overestimate the zero proportion by about 24\%. 

For the purpose of pricing rainfall derivatives that depend on accumulated rainfall amounts, it is not particularly important to precisely match the zero proportions of the data. For other applications, a different approach for estimating the parameters of $L$ may be more appropriate. One alternative would be to use a simulation-based generalised method of moments, with one of the moment conditions specifying that the proportion of implied zero values in empirical and simulated data match.   

\begin{table}
\centering
\caption{Proportion of implied zero values in empirical data and corresponding simulation average. 
\label{zero}}
\begin{tabular}{llll}
\hline
Location & Time scale & Type & Implied zero proportion (\%) \\
\hline
Heathrow & Hourly & Simulation average& 90.39 \\
& & Data &  91.27\\
\hline
Detroit & Daily & Simulation average & 59.52 \\ 
& & Data &  48.14\\
\hline
\end{tabular}
\end{table}

\subsection{Monthly fit}
In this subsection we consider the fit of the model on a monthly basis, which is especially relevant for the pricing application in Section \ref{pricing}. Although the current derivatives at the CME only consider total monthly accumulated rainfall, rainfall derivatives are sold over-the-counter, i.e.~traded directly between two parties, so they can be tailored to specific needs. Thus one can consider derivatives that depend on daily accumulations or other relevant quantities. Hence it is important to have a flexible modelling framework which allows for adjustments to different time scales.

\begin{figure}[tbp]
  \centering
  \includegraphics[width=0.7\textwidth]{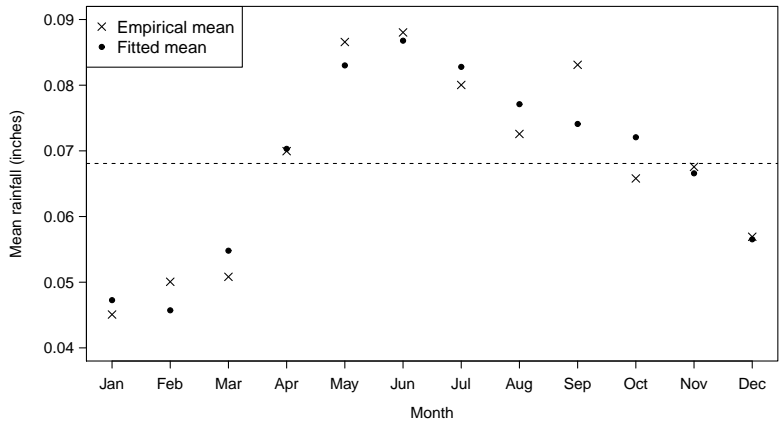}
  \caption{Empirical and fitted monthly means for daily Detroit rainfall, stipled line shows overall mean.}
  \label{dmean}
\end{figure}

Figure \ref{dmean} shows the fitted and empirical monthly means for the Detroit rainfall. Because the mean of $\triangle Y$ is constant, this plot gives a measure of the fit of the truncated Fourier series used for the seasonality function $S$. The plot also demonstrates the fit to the monthly totals that form the basis of the current CME derivatives. We see that the fitted seasonality matches the yearly trend quite well, except for the months of September and October, where the empirical means deviate from the smooth curve.

\begin{figure}[tbp]
  \centering
  \includegraphics[width=0.9\textwidth]{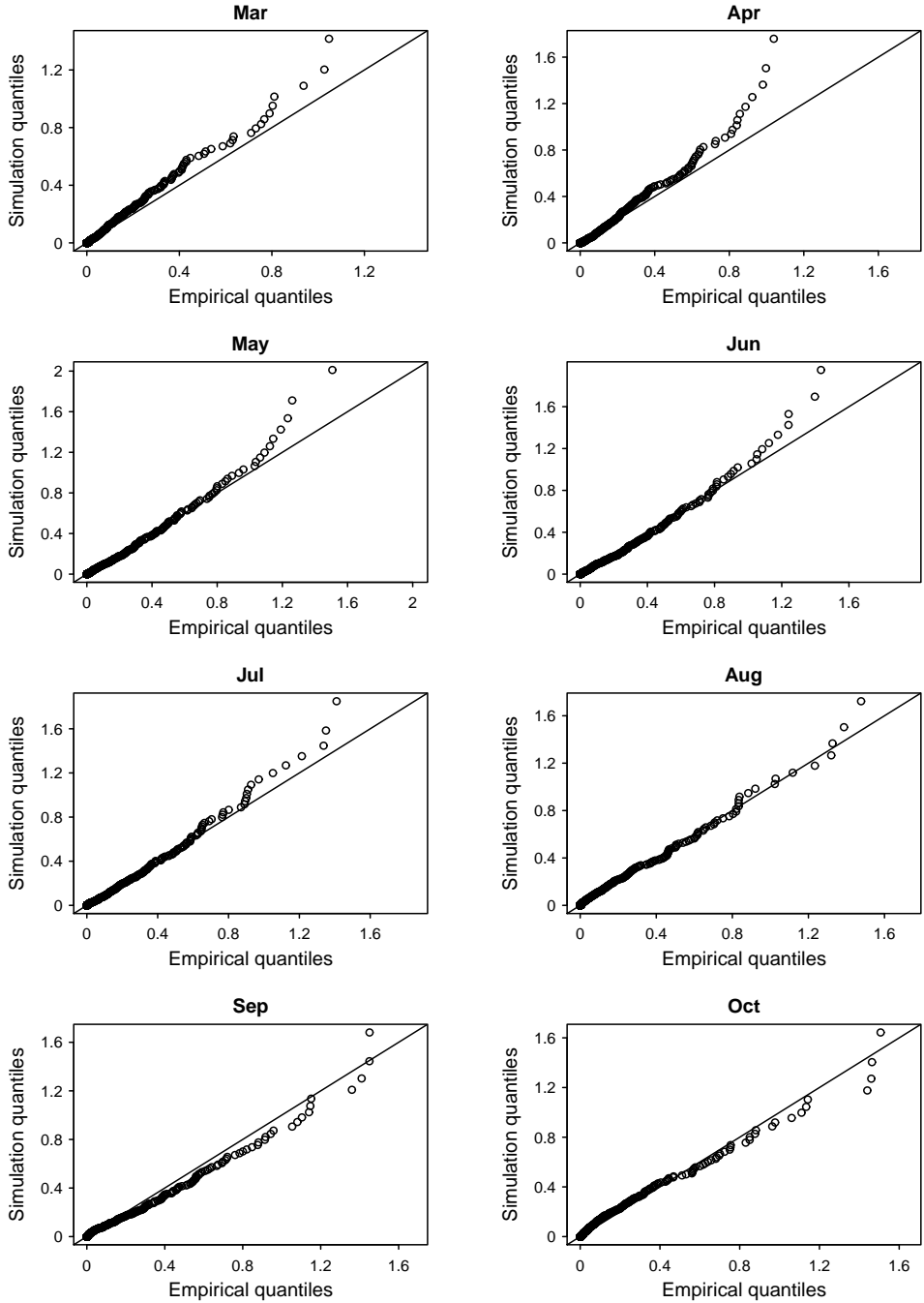}
  \caption{QQ-plots comparing simulated and empirical daily Detroit rainfall for months included in current CME rainfall derivatives.}
  \label{dqq}
\end{figure}

Figure \ref{dqq} shows plots comparing the quantiles of the empirical and simulated Detroit daily rainfall for the months from March to October, which are the months considered for the CME rainfall derivatives currently on offer. As mentioned above, derivatives could be made to depend on rainfall accumulated over different time scales, thus it is reasonable to consider the fit to the finest available time scale, i.e.~daily data. The simulation quantiles are based on combined data from the 500 simulations used for the frequency plots in Section \ref{marginal}. Similar to the overall QQ-plot in Figure \ref{qqplots}, the overall fit is acceptable, especially in the lower end of the data range (the 99.9 percentile of the empirical daily rainfall is $1.46$). 

\section{Rainfall derivative pricing} \label{pricing}
In this section we calculate prices for rainfall futures contracts based on the daily rainfall model presented in this paper. This is done by first deriving the characteristic function of a general L\'evy-driven stochastic process under a risk-neutral measure. More precisely, we will work with the risk-neutral measure induced by the Esscher transform. In the following section we work on a complete probability space $(\Omega,\mathcal{F},P)$. 

\subsection{Pricing methodology} \label{method}
Classical asset pricing theory is based on the assumption of a complete market, where the risk associated with any derivative can be completely hedged against by replicating the derivative through a portfolio that includes holdings of the underlying asset. Then the derivative has a unique fair price equal to that of the replicating portfolio, and we say that the market is complete. This price can also be specified as the expected final payoff of the derivative under an equivalent measure $Q$, called the risk-neutral measure. Under this measure the discounted price processes of all tradeable assets are martingales.  

For rainfall derivatives the underlying ``asset'' is an index $(I(t))_{t \geq 0}$ measuring accumulated rainfall, which cannot be directly traded, and so the hedging argument cannot be applied. Thus the market for rainfall derivatives is incomplete, meaning that there is no unique fair price of the derivative. Hence there exist many possible choices of equivalent probability measures. In the present paper we construct one such measure by using the Esscher transform on the underlying rainfall process, which we specify through our L\'evy-driven rainfall model.  

The Esscher transform is a generalised Girsanov transform for jump processes; it was first introduced by \cite{Esscher1932} as a change of probability measure, and \cite{Gerber1994} generalised the transform to stochastic processes driven by a L\'evy process. As shown in \cite{Esche2005}, the Esscher transform preserves the L\'evy properties of the process to be transformed. This property makes it a natural choice for constructing a risk-neutral measure when the underlying is driven by a L\'evy process, and contributes to achieving analytical tractability.   

In the following we consider a finite time horizon $T<\infty$, and assume all derivatives expire before that time. We also consider a L\'evy subordinator $(L(t))_{t \geq 0}$ (assumed to be c\`adl\`ag), and extend $L$ to a two-sided L\'evy process $(L^*(t))_{t \in \mathbb{R}}$ by defining the process $\hat{L}$ to be an independent 
(c\`adl\`ag) copy of $L$ such that $L$ and $\hat{L}$ have the same characteristic triplet, and letting 
\begin{align} \label{twosided}
L^*(t)= \begin{cases}
   L(t), & \mbox{for } t \geq 0 \\
   -\hat{L}(-(t-)),  & \mbox{for } t < 0,
  \end{cases}
\end{align}
which makes $L^*$ c\`adl\`ag. In the following we will take $L$ to mean the two-sided process $L^*$ in order to simplify notation. We define the so-called \textit{increment filtration} \citep{OConnor2014} by
\begin{align} \label{increment}
\mathcal{F}_t=\sigma(L_u-L_s: -\infty<s<u \leq t), \; t \in \mathbb{R},
\end{align}
so that $(L(t))_{t \in \mathbb{R}}$ is a L\'evy process in this filtration. 

We will use the generalised version of the Esscher transform for a L\'evy process $(L(t))_{t \in \mathbb{R}}$  with filtration $(\mathcal{F}_t)_{t \in \mathbb{R}}$ as above, which is defined by giving the Radon-Nikodym derivative 
\begin{equation} \label{Esscher}
\left. \frac{dQ}{dP}\right|_{\mathcal{F}_t}=Z(t)=\frac{\exp\left\{\int_0^t  \theta(s) dL(s) \right\}}{E\left[\exp\{\int_0^t  \theta(s) dL(s) \} \right]},
\end{equation}
where $\theta(s)$ is a time-dependent parameter, as opposed to the standard transform where it is constant. This parameter can be interpreted as a measure of risk-aversion, called the \emph{market price of risk} (MPR), and is used to calibrate $Q$ such that theoretical and observed market prices match. Specifically, the investor selling a derivative at time $t$ will have to pay an amount given by the payoff function of the index $I$ at the time of maturity $\tau$. This amount is determined by the jumps of the driving process $L$ in the future time interval $[t,\tau]$. Thus the investor is exposed to risk from these jumps, and the Esscher transform reflects the corresponding risk premium by exponentially tilting the jump measure.   

Having defined $Q$ via the Esscher transform, we find derivative prices by taking expected values of payoffs at maturity under $Q$, conditional on the information known at the current time, similar to the complete market case. For simplicity we assume a zero interest rate. Then for a rainfall index $I$ adapted to the filtration $(\mathcal{F}_t)_{t \geq 0}$, we get that the futures price process given by
\[f_\tau(t)=E_Q[I(\tau)|\mathcal{F}_t],\] 
will be a $Q$-martingale by construction (provided it is integrable), which is required since the derivative contract is itself a tradeable asset. However, since the market is incomplete we do not require the underlying rainfall index process $(I(t))_{t \geq 0}$ to be a $Q$-martingale, since it cannot be directly traded. 

\subsection{Esscher transform for integrated moving average processes}
In this subsection we show the result of applying the Esscher transform to the class of moving average processes, which includes our rainfall model as a special case. 

Consider a two-sided L\'evy subordinator $(L(t))_{t \in \mathbb{R}}$ with associated filtration $(\mathcal{F}_t)_{t \in \mathbb{R}}$, as defined in \eqref{twosided} and \eqref{increment}. We now define the stochastic process $X(t)$ by
\begin{align*}
X(s)=\int_{-\infty}^s h(s-v) \, dL(v)=\underbrace{\int_{-\infty}^0 h(s-v) \, dL(v)}_{:=\tilde{A}(s)} +\int_0^s h(s-v) \, dL(v),
\end{align*}
where $h: \mathbb{R}_+ \rightarrow \mathbb{R}_+$ is a left-continuous, square integrable deterministic function such that $h \in L^1$. The resulting process $X$ is strictly stationary, and is called a \emph{moving average process} \citep{Applebaum}. A moving average process can be seen as a general form of the Ornstein-Uhlenbeck process, for the OU equation given by \eqref{xeqn} we have $h(s)=\exp\{ -\lambda s\}$, with $\tilde{A}(s)= e^{-\lambda s} X(0)$. 

If we now integrate $X$ over the interval $[0,t]$ and exchange the order of integration by using the stochastic Fubini theorem, we get the integrated moving average process, which is similar to the primary rainfall model given in \eqref{yeqn}:
\begin{align} \label{intlevy} 
\begin{split}
Y(t)&=\int_0^t X(s) \, ds = \underbrace{\int_{-\infty}^0 \tilde{g}(t,v) \, dL(v)}_{:=A(0,t)}+\int_0^t g(t,v)\, dL(v), \\
g(t,v)&=\int_v^t h(s-v) \, ds, \quad \tilde{g}(t,v)=\int_0^t h(s-v) \, ds, 
\end{split}
\end{align}
with $A(0,t) \in \mathcal{F}_0 \, \forall t$.  

We now want to calculate the characteristic function of the process $Y$ under the probability measure $Q$ specified by the Esscher transform defined in \eqref{Esscher}. In order to ensure that the Radon-Nikodym derivative $Z$ is well-defined, we assume that $L$ satisfies the exponential moment condition, which states that there exists a constant $k>0$ such that
\begin{align} \label{momentcond}
E[\exp(kL(t))]<\infty,
\end{align}
for $t<T$, where $T$ is our time horizon. For the particular case given by our rainfall model, $L$ is the Hougaard L\'evy process $L(\mu,\rho,\kappa)$, which has exponential moments for $k < \mu^{1-\kappa}/(\rho(\kappa-1))$.

As discussed in Subsection \ref{method}, derivative prices are calculated in terms of the expected payoff at maturity under the measure $Q$, conditional on the current information $(\mathcal{F}_t)$. We want to find prices for a general payoff function $f(\mbox{Ind}(\tau_1,\tau_2))$, where $\mbox{Ind}(\tau_1,\tau_2)=Y(\tau_2)-Y(\tau_1)$ is the index measuring accumulated rainfall in the interval $[\tau_1,\tau_2]$. We follow \cite{Benth2013} in using Fourier methods for these calculations, where we define the Fourier transform and its inverse by
\begin{align} \label{ft}
\begin{split}
\hat{f}(y)&=  \int_{\mathbb{R}} f(x) e^{-ixy} \, dx, \\
f(x)& = \frac{1}{2 \pi} \int_{\mathbb{R}} \hat{f}(y) e^{ixy} \, dy,
\end{split}
\end{align}
assuming $f, \hat{f} \in L^1(\mathbb{R})$. 
To account for the case when $f$ is not integrable, we consider the function $f_{\delta}(x)=e^{-\delta x} f(x)$, and assume there exists a $\delta \geq 0$ such that $f_\delta$ is integrable. We then have the following result specifying the risk-neutral expectation of  $f(Y(\tau_2)-Y(\tau_1))$.  

\begin{proposition} \label{pricingprop}
Let $f$ be a payoff function such that $\hat{f}_\delta \in L^1$, where $\hat{f}$ is the Fourier transform of $f$ as defined in \eqref{ft}. Assume that
\[\sup_{v \in [t,\tau_2]} \left(\delta|g(\tau_2,v)-g(\tau_1,v)|+|\theta(v)| \right)<k,\]
for $k$ given by the exponential moment condition. Then, when $Y$ is an integrated moving average process as specified in \eqref{intlevy}, and $Q$ corresponds to the Esscher transform, we have that
\begin{align} \label{qformula}
\begin{split}
&E_Q \left(f\left(Y(\tau_2)-Y(\tau_1)\right) |\mathcal{F}_t  \right) \\
&=\frac{1}{2 \pi} \int_{\mathbb{R}} \hat{f}_\delta(\xi)  \exp\left\{(\delta+i\xi)\left(A(0,\tau_2)-A(0,\tau_1) + \int_0^t [g(\tau_2,v)-g(\tau_1,v)] \, dL(v) \right) \right. \\
&+ \left.  \int_t^{\tau_1} \psi_\theta(v,(\delta+i \xi)[g(\tau_2,v)-g(\tau_1,v)]) \, dv +\int_{\tau_1}^{\tau_2} \psi_{\theta}(v,(\delta+i \xi)g(\tau_2,v)) \, dv  \right\} \, d\xi,   
\end{split}
\end{align}
for $t < \tau_1 <  \tau_2<T$, where we define
\begin{align*}
\psi_\theta(s,\gamma c(s)):=\int_{\mathbb{R}_+} e^{\theta(s)y}(e^{(i \gamma c(s)y}-1) \nu(dy),
\end{align*}
for a complex variable $\gamma$ and real-valued function $c$, where $\nu(\cdot)$ is the L\'evy measure of $L$. 
\end{proposition}
The proof of Proposition \ref{pricingprop} is given in the Appendix. 

We stress that the resulting derivative price \emph{does} depend on $t,\tau_1,\tau_2$, and as such our model constitutes an important generalisation of the independent increment model considered in ~\cite{Benth2013}. In their setup, the resulting futures prices did not exhibit any dynamics in $t$, which is a serious restriction in practical applications. 

\subsection{Pricing illustration}
We now illustrate how to calculate the prices for futures written on the Detroit rainfall by using the model presented in Section \ref{model}, with the estimated parameters given in Table \ref{dpars}. For rainfall futures the payoff equals the index $Y(\tau_2)-Y(\tau_1)$, hence such contracts are also called swap contracts. 

Recall that for the Detroit rainfall we used an integrated CARMA model of order $p=1$, which reduces to the integrated OU model. This model can be written in the general form given in \eqref{intlevy} above, with
\[g(t,s)=\frac{1-e^{-\lambda(t-s)}}{\lambda}.\] 

Letting $f(x)=e^{iux}$, we obtain the characteristic function $E_Q(e^{iu(Y(\tau_2)-Y(\tau_1))}|\mathcal{F}_t)$ as the exponential term in \eqref{qformula} evaluated at $\delta=0$ and $\xi=u$. Taking derivatives with respect to $u$ and evaluating at $u=0$ gives 
\begin{align*} 
&E_Q\left[Y(\tau_2)-Y(\tau_1)\left. \right|\mathcal{F}_t\right]=A(0,\tau_2)-A(0,\tau_1)+\left(\frac{e^{-\lambda \tau_1}-e^{-\lambda \tau_2}}{\lambda}\right)\int_0^t e^{\lambda v} \, dL(v) \\
&+\int_t^{\tau_1} -i\psi_\theta '(v,0) \left(\frac{e^{-\lambda \tau_1}-e^{-\lambda \tau_2}}{\lambda}\right) \, dv+\int_{\tau_1}^{\tau_2} -i\psi_\theta'(v,0) \left(\frac{1-e^{-\lambda(\tau_2-v)}}{\lambda}\right) \, dv ,
\end{align*}
where $\psi_\theta'$ denotes the derivative with respect to the second argument. The explicit form of $\psi_\theta$ for this model can be found analytically in terms of the parameters of the Hougaard process, and is specified in the Appendix. 
When $L$ is a Hougaard process the exponential moment condition \eqref{momentcond} restricts the values of $\theta(v)$ to be below $\mu^{1-\kappa}/(\rho(\kappa-1)$, which equals $0.042$ when using the estimated parameters. Note that this restriction does not affect the range of the prices, as the price explodes when $\theta(v)$ approaches the upper limit. 

For our rainfall data we need to evaluate this expression based on the discrete observations $(Y(t_i))$, meaning that we do not observe $L$ or $A(0,\tau_i)$ directly. To obtain an explicit value for the price we approximate these terms by their expected value. It can be shown that for larger values of $\tau_1-t$ the unobserved terms are negligible compared to the last two terms; hence the mean approximation does not significantly affect the value of the price. 

We calculated prices for monthly rainfall contracts in 2011 for a time $t$ corresponding to the 31st of December 2010. The final price equals 
\[S_{m_i}(t)E_Q[Y(\tau_2)-Y(\tau_1) \left. \right|\mathcal{F}_t],\] 
where $S_{m_i}(t)$ is the monthly average of the seasonality function $S(t)$. Table \ref{tab:DetroitRainfallFutureMonthlyPrices} shows a range of the resulting prices corresponding to different values of the parameter $\theta$, representing the risk premium. 

The first row of the table shows market prices reported by the CME, and we see that by adjusting $\theta$ we can calibrate the prices obtained from the model to match the market price. The values of $\theta$ resulting from this calibration are shown in Table \ref{est_theta}. We remark that rainfall derivatives are a very recent addition to the CME portfolio, and their trading volume is currently close to zero, thus the reported CME prices for 2011 do not accurately reflect the market value of these products at the current time. Hence the corresponding estimates of $\theta$ for the specific 2011 prices may also differ from their true value. These values do however demonstrate how the rainfall model and associated pricing methodology provides a unified and flexible framework for studying the market view of the risk associated with rainfall.     

\begin{table}[tbp]
\centering
\caption{Prices of monthly rainfall contracts for Detroit.}
\label{tab:DetroitRainfallFutureMonthlyPrices}
\small
\tabcolsep=0.15cm
\begin{tabular}{l|r@{}l|r@{}lr@{}lr@{}lr@{}lr@{}lr@{}lr@{}lr@{}l}
\hline
&\multicolumn{2}{c|}{$\theta$} & \multicolumn{2}{r}{Mar 11} & \multicolumn{2}{r}{Apr 11}& \multicolumn{2}{r}{May 11}&\multicolumn{2}{r}{Jun 11}& \multicolumn{2}{r}{Jul 11}&\multicolumn{2}{r}{Aug 11}& \multicolumn{2}{r}{Sep 11}& \multicolumn{2}{r}{Oct 11}\\ 
\hline
CME price && &4&.2 &4&.4 &3&.2 &5&.0 &4&.5 &4&.3 &4&.2 &4&.6 \\
\hline
Model price & -0&.01 &1&.20 & 1&.50 & 1&.83 & 1&.85 & 1&.83 & 1&.70 & 1&.58 & 1&.59 \\      
& 0&.00 & 1&.69 & 2&.10 & 2&.57 & 2&.60 & 2&.57 & 2&.39 & 2&.22 & 2&.24  \\
&0&.01 &2&.61 & 3&.24 & 3&.96 & 4&.01 & 3&.96 & 3&.69 & 3&.43 & 3&.45  \\ 
&0&.02 & 4&.72 & 5&.86 & 7&.15 & 7&.25 & 7&.16 & 6&.67 & 6&.19 & 6&.23  \\
&0&.03 & 12&.12 & 15&.04 & 18&.38 & 18&.63 & 18&.39 & 17&.12 & 15&.91 & 16&.01\\
&0&.04&153&.69 & 190&.80 & 233&.11 & 236&.27 & 233&.21 & 217&.19 & 201&.83 & 203&.13\\
\hline
\end{tabular}
\end{table}

\begin{table}[tbp]
\centering
\caption{Estimated values of parameter $\theta$ based on CME prices.}
\label{est_theta}
\small
\begin{tabular}{cllllllll}
\hline
Month & Mar 11& Apr 11& May 11& Jun 11& Jul 11 & Aug 11& Sep 11& Oct 11\\ 
$\theta$ & 0.0183 & 0.0156 & 0.0054 & 0.0142 & 0.0125 &  0.0130 & 0.01314 & 0.0142 \\
\hline
\end{tabular}
\end{table}
   
\section{Conclusion}
We have introduced a new class of continuous-time stochastic processes, driven by the Hougaard L\'evy process, and shown how it can be used to construct a parsimonious and analytically tractable model for rainfall. By generalising the Ornstein-Uhlenbeck process representing rainfall intensity to a continuous-time ARMA (CARMA) process, we obtain a model with a very flexible autocorrelation structure. We presented a general fitting method for this class which exploits a correspondence between integrated CARMA and ARMA processes. 

We showed that the model fits the marginal distribution of the rainfall very well on both hourly and daily time scales.  In particular, the marginal fit for daily rainfall is better than that of the standard model described in \cite{Wilks1998}, and we also obtain an adequate fit to the mean and marginal distribution on a month-by-month basis. 

By virtue of the CARMA generalisation, the extended model manages to accurately reproduce the autocorrelation structure of the observed rainfall, a characterising feature of the process which becomes increasingly significant for smaller time scales. 

The last part of the paper gives a result specifying the risk-neutral expectation of a function of the rainfall process, which can be used for pricing general derivatives written on a precipitation index. To construct a risk-neutral measure we use the Esscher transform, with a time-dependent parameter representing the risk premium. We state the result for a general moving average process, a class which includes our model as a special case. The pricing methodology constitutes an important generalisation of the independent increment model considered in \cite{Benth2013}, which does not allow for price dynamics of derivatives. 

We illustrated the pricing method by calculating futures prices based on empirical daily rainfall data from Detroit, and showed how they can be calibrated to observed prices. Given a sufficient amount of price data, one can study the properties of the market price of risk (MPR) that is implied by this calibration. \cite{Hardle2012} discuss the market price of weather risk, focusing on temperature derivatives, and provides an example of modelling the  MPR of temperature derivatives as a deterministic function of the seasonal temperature variation. As the market for rainfall derivatives matures, one could conduct a similar study for the MPR of rainfall derivatives.

\paragraph{Acknowledgements.}
We thank the Associate Editor and two anonymous referees for constructive suggestions that led to significant improvements of the paper. R. C. Noven gratefully acknowledges financial support from the Grantham Institute for Climate Change, Imperial College London. We thank the UK Meteorological Office and the British Atmospheric Data Centre for providing the data used. 

\section{Appendix} \label{proofs}
In the following we present the proofs of our theoretical results. 

First we quote a result \citep[Lemma 15.1]{Cont2004} which will be used repeatedly in the following:  

\begin{lemma} \label{intlem}
Let $f: [0,T]\rightarrow \mathbb{R}$ be a left-continuous function and $L(t)$ a L\'evy process. Then 
\begin{align*}
E\left[\exp \left\{\int_0^t if(s) dL(s) \right\} \right]=\exp \left\{\int_0^t \psi(f(s)) ds, \right\}
\end{align*}
where $\psi(t)$ is the characteristic exponent of $L$, given by
\[\exp\{\psi(u)\}=E[e^{iuL(1)}].\]
\end{lemma}

\textbf{Characteristic function of $\triangle Y$}.
The characteristic function of $\triangle Y$ is given by
\[\varphi_{\triangle Y}=\exp\left\{\int_0^{\infty}\psi(ug_1(s))\, ds+\int_0^1 \psi(ug_2(s)) \, ds\right\},\]
where 
\begin{align*}
g_1(s)&=\sum_{k=1}^p \frac{e^{-\lambda_k s}-e^{-\lambda_k(1+s)}}{\lambda_k}, \\
g_2(s)&=\sum_{k=1}^p \frac{1-e^{-\lambda_k(1-s)}}{\lambda_k}.
\end{align*}

This follows immediately from applying Lemma \ref{intlem} to the expression given in \eqref{int_form}, and noting that because $\triangle Y$ is stationary we can set $t_{i-1}=t_0=0$, causing the second integral to vanish. 

\vspace{10pt}

\textbf{Proof of Proposition \ref{pricingprop}.}
By construction of $f_\delta$, we have that
\[f(x)=\frac{1}{2 \pi} \int_{\mathbb{R}} \hat{f}_\delta(\xi) e^{(\delta+i\xi)x} \, d\xi,\]
and hence by the Fubini theorem, it follows that
\[E_Q[\left. f(Y(\tau_2)-Y(\tau_1)) \right| \cf_t]=\frac{1}{2 \pi} \int_{\mathbb{R}} \hat{f}_\delta(\xi) E_Q[\left. e^{(\delta+i\xi)(Y(\tau_2)-Y(\tau_1))} \right| \cf_t] \, d\xi,\]
similar to the proof of Proposition 8.4 in \cite{Benth2013}. 

We now calculate the expectation involving the integrated moving average process $Y$. To this end, we first split the integrals in the expression for $Y$ as follows: 
\begin{align} \label{priceparts}
\begin{split} 
&E_Q \left. \left[\exp \left\{(\delta+i \xi)(Y(\tau_2)-Y(\tau_1)) \right\} \right|\, \mathcal{F}_t  \right]  \\
=& \exp \left\{(\delta+i \xi)\left(A(0,\tau_2)-A(0,\tau_1) + \int_0^t \left[g(\tau_2,v)-g(\tau_1,v) \right] \, dL(v) \right)\right\}  \\
\times & \underbrace{E_Q \left[ \left. \exp \left\{(\delta+i \xi) \int_t^{\tau_1}\left[g(\tau_2,v)-g(\tau_1,v) \right] \, dL(v) \right\} \right| \mathcal{F}_t \right]}_{\mathbf{(A)}} \\ 
&\times  \underbrace{E_Q \left[ \left. \exp \left\{(\delta+i \xi) \int_{\tau_1}^{\tau_2} g(\tau_2,v) \, dL(v) \right\} \right| \mathcal{F}_t \right]}_{\mathbf{(B)}}. 
\end{split}
\end{align}
By the abstract Bayes formula (see e.g \citealp{Oksendal2000}), for the measure $Q$ such that $dQ/dP|_{\mathcal{F}_t}=Z(t)$, with $X$ being $\mathcal{F}_\tau$-measurable and $t<\tau$, we have that 
\[E_Q(X|\cf_t)=E\left(\left. X \frac{Z(\tau)}{Z(t)} \right|\cf_t \right).\]

Recall that we are working with the Esscher transform, so we have
\[Z(t)=\frac{\exp\left\{\int_0^t \theta(v) \, dL(v) \right\}}{E\left[\exp\left\{\int_0^t \theta(v) \, dL(v) \right\} \right]}.\]
Applying the Esscher transform then gives
\begin{align*}
\mathbf{(A)} = \, &E \left[ \left. \exp \left\{\int_t^{\tau_1}  (\delta+i \xi) \left[g(\tau_2,v)-g(\tau_1,v) \right] \, dL(v) \right\} \frac{Z(\tau_1)}{Z(t)}\right| \mathcal{F}_t \right] \\
=&E \left[ \exp \left\{\int_t^{\tau_1} \left((\delta+i \xi) \left[g(\tau_2,v)-g(\tau_1,v) \right] +\theta(v) \right) \, dL(v) \right\} \right]
\times \frac{E\left[\exp \left\{\int_0^t \theta(v) \, dL(v) \right\} \right]}{E \left[\exp \left\{\int_0^{\tau_1} \theta(v) \, dL(v) \right\} \right]},
\end{align*}
where we get an unconditional expectation due to the independent increments of $L$. 

We can extend Lemma \ref{intlem} to complex-valued functions to get 
\begin{align} \label{extlemma}
E\left[\exp \left\{\int_0^t (a(v)+ib(v)) \, dL(v) \right\} \right]=\exp \left\{\int_0^t \psi_1(-ia(v)+b(v)) \, dv \right\},
\end{align}
where the term on the RHS equals
\begin{align*}
\exp \left\{\int_0^t \ln E \left(\exp\{a(v)+ib(v)\}L(1) \right) \,  dv \right\},
\end{align*}
and we have that
\begin{align*}
\left|E \left(e^{(a(v)+ib(v))L(1)} \right) \right| \leq E(e^{a(v)L(1)}),
\end{align*}
and so if $\sup_v a(v)<k$, then the last term is bounded by the exponential moment condition given in \eqref{momentcond}. 

Applying \eqref{extlemma} to the terms in $\mathbf{(A)}$ gives
\begin{align*} 
\mathbf{(A)}
&=\exp \left\{\int_t^{\tau_1} \psi \Big(\xi \left[g(\tau_2,v)-g(\tau_1,v) \right] -i\left[\theta(v)+\delta \left(g(\tau_2,v)-g(\tau_1,v) \right) \right] \Big) - \psi \left(-i\theta(v) \right) \, dv\right\},
\end{align*}
where the requirement $\sup_{v \in [t,\tau_2]} \left(|\theta(v)|+\delta|g(\tau_2,v)-g(\tau_1,v)| \right)<k$ ensures that the terms in the above equation are well-defined.
 
Now we consider the L\'evy-Khintchine formula for subordinators, which takes the form
\begin{align*} 
E(e^{iuL(t)})=\exp \left\{t \psi(u) \right\}=\exp \left\{t \int_{\mathbb{R}_+} (e^{iuy}-1) \nu(dy) \right\}, 
\end{align*} 
where $\nu(\cdot)$ is the L\'evy measure associated with $L$. We can analytically continue this formula to complex arguments \citep[p. 338]{Applebaum}, and so we get that
\begin{align*}
\mathbf{(A)}
&= \exp\left\{\int_t^{\tau_1} \int_{\mathbb{R}_+} e^{\theta(v)y}(e^{(\delta+i \xi) \left[g(\tau_2,v)-g(\tau_1,v) \right]y}-1) \nu(dy) \, dv \right\}. 
\end{align*}
We note that this expression can also be written as
\begin{align*}
E\left[\exp\left\{\int_t^{\tau_1} (\delta+i \xi) \left[ g(\tau_2,v)-g(\tau_1,v) \right] \, dL_Q(v) \right\} \right],
\end{align*}
where $L_Q(v)$ is now a non-stationary stochastic process with jump measure depending on time, namely
\begin{align*}
\nu_\theta(dv,dy)=e^{\theta(v)y} \nu(dy) dv.
\end{align*} 

Thus we see that conditioning with respect to the measure $Q$ has the effect of exponentially tilting the jump measure of $L$ at time $v$ according to $\theta(v)$, so the jumps of $L$ at times $v$ will be weighted more or less in the expectation depending on the sign of $\theta(v)$. 

Now defining 
\begin{align} \label{psitheta}
\psi_\theta(v,\gamma c(\cdot)):=\int_{\mathbb{R}_+} e^{\theta(v)y} (e^{\gamma c(v)y}-1) \nu(dy),
\end{align}
we get that
\begin{align*}
\mathbf{(A)}=\exp \left\{\int_t^{\tau_1} \psi_\theta \Big(v,(\delta+i \xi)\left[g(\tau_2,v)-g(\tau_1,v) \right] \Big) \, dv \right\};
\end{align*}
and by similar arguments
\begin{align*}
\mathbf{(B)}=\exp \left\{\int_{\tau_1}^{\tau_2} \psi_\theta \Big(v,(\delta+i \xi)g(\tau_2,v) \Big) \, dv \right\}.
\end{align*}
Substituting these expressions into \eqref{priceparts} then gives the result. 

\vspace{10pt}

\textbf{Hougaard process.}
The Hougaard process has L\'evy measure given by \citep{Grigelionis2011} 
\[\nu(dy)= \left(\rho^{\frac{1}{\kappa-1}} \Gamma \left(\frac{\kappa}{\kappa-1} \right) (\kappa-1)^{\kappa/(\kappa-1)} \right)^{-1} y^{\frac{3-2\kappa}{\kappa-1}} \exp \left\{- \frac{\mu^{1-\kappa}}{\rho (\kappa-1)} y \right\} dy,\]
in terms of the Tweedie parameterisation. We also have that when $L$ is the Hougaard process, the function $\psi_\theta$ defined in \eqref{psitheta} takes the form
\[\psi_\theta(v,\gamma c(v))=\frac{\mu^{2-\kappa}}{\rho(2-\kappa)} \left[\left(1-\frac{\rho(\kappa-1)\left(i \gamma c(v)+\theta(s)\right)}{\mu^{1-\kappa}} \right)^{(\kappa-2)/(\kappa-1)}-1 \right].\]

\end{document}